\documentstyle[12pt]{article}

\newcommand{\no}{\noindent}

\newcommand{\be}{\begin{equation}} 
\newcommand{\bea}{\begin{eqnarray}}
\newcommand{\ee}{\end{equation}}
\newcommand{\eea}{\end{eqnarray}}

\def\G{\Gamma}
\def\S{{\bf S}}
\def\R{{\bf R}}
\def\Z{{\bf Z}}
\def\C{{\bf C}}
\def\G{\Gamma}
\def\N{{\cal N}}

\begin{document}

\flushright {hep-th/0203256}
\vskip .5cm
\Huge
\centerline{On the Quantum Moduli Space}
\centerline{of $M$-Theory Compactifications} 
\vskip 2cm
\large
\centerline{Tamar Friedmann\footnote{tamarf@feynman.princeton.edu}}
\vskip .2cm
\normalsize
\centerline{\it Joseph Henry Laboratories, Princeton University}
\centerline{\it Princeton, NJ 08544, USA}
\vskip 3cm
\renewcommand{\baselinestretch}{1.3}
\abstract{}
We study the moduli space of $M$-theories compactified on $G_2$ manifolds which are asymptotic to a cone over quotients of $\S^3\times \S^3$. We show that the moduli space is composed of several components, each of which interpolates smoothly among various classical limits corresponding to low energy gauge theories with a given number of massless $U(1)$ factors. Each component smoothly interpolates among supersymmetric gauge theories with different gauge groups.

\newpage
\renewcommand{\baselinestretch}{1.3}
\normalsize
\section{Introduction}

The study of $M$-theory compactifications on seven dimensional manifolds $X$ of $G_2$ holonomy has been motivated by the fact that such compactifications result in unbroken supersymmetry in four dimensions. The properties of the compactification manifold $X$ determine the particle spectrum of the corresponding four dimensional theory. It has been shown in recent years that compactifications on singular manifolds can result in low energy physics containing interesting massless spectra. Specifically, 
certain singular $G_2$ manifolds give rise to $\N =1$ supersymmetric gauge theories at low energies, as shown for example in \cite{Ach1, Ach2,AMV}. There, $X$ was taken to be asymptotic to a quotient of a cone on $\S^3\times \S^3$, and the singularities of $X$ took the form of families of $ADE$ singularities giving $ADE$ gauge theories at low energies.  

Subsequently, the quantum moduli space of $M$-theories on $G_2$ manifolds $X$ which are asymptotic to a cone on $\S^3\times \S^3$ or quotients thereof has been studied in \cite{AW}. It was shown that the moduli space is a Riemann surface of genus zero, which interpolates smoothly between different semiclassical spacetimes. 

The purpose of this paper is to generalize the construction of \cite{AW} to other quotients of $\S^3\times \S^3$ and obtain the moduli spaces for those as well. Our quotients contain those in \cite{AW} as special cases. We propose that the moduli space for our quotients consists of several branches classified according to the number of massless $U(1)$ factors that appear in the low energy gauge theories corresponding to semiclassical points. Each branch of the moduli space interpolates smoothly between the different semiclassical points appearing on it; hence, we get smooth interpolation between supersymmetric gauge theories with different gauge groups.

This paper is organized as follows: in Section \ref{Y} we review the $M$-theory dynamics on the cone on $Y=\S^3\times \S ^3$ given in \cite{AW}. In Section \ref{quotients}, we describe quotients of this cone by discrete groups of the form $\Gamma = \G_1\times \G_2\times \G_3$ where the $\G_i$ are $ADE$ subgroups of $SU(2)$; these $ADE$ groups must be chosen carefully in order to obtain known low energy gauge theories from the compactification. In Section \ref{space}, we turn to the description of the moduli space ${\cal N}_\Gamma$ of $M$-theories on these quotients, beginning with the classical moduli space and concluding with the quantum moduli space. 

While this paper was being completed, we received \cite{Oz} which has overlap with the case where $\G_3$ is trivial (or $r=1$ in our notation of Section \ref{quotients}).


\section{Dynamics of $M$-theory on the Cone over $\S ^3 \times \S^3$}\label{Y}

In this section, we review the $M$-theory dynamics on a manifold $X$ of $G_2$ holonomy which is asymptotic at infinity to a cone over $Y= \S^3\times \S^3$ \cite{AW}. The manifold $Y$ can be described as a homogeneous space $Y=SU(2)^3/SU(2)$, where the equivalence relation is $(g_1, g_2, g_3)\sim (g_1h,g_2h,g_3h)$, $g_i,h\in SU(2)$. Viewed this way, this manifold has $SU(2)^3$ symmetry via left action on each of the three factors in $Y$, as well as a ``triality'' symmetry $S_3$ permuting the three factors. Up to scaling, there is a unique metric with such symmetries given by
\be d\Omega ^2 = da^2+db^2+dc^2 ,\ee 
where $a,b,c\in SU(2)$, $da^2= -\mbox{Tr}(a^{-1}da)^2$, the trace is taken in the fundamental representation of $SU(2)$, and $a,b,c$ are related to $g_1,g_2,g_3$ by $a=g_2g_3^{-1}$ and cyclic permutations thereof.

The metric for a cone on $Y$ is
\be \label{cone} ds^2=dr^2+r^2d\Omega ^2 ,\ee
where $d\Omega ^2$ is the metric on $Y$. Such a cone can be constructed by filling in one of the three $SU(2)\sim \S^3$ factors of $Y$ to a ball. We denote the manifold obtained by filling in a given $g_i$ by $X_i$. The metric on a manifold $X$, asymptotic to $X_i$ at infinity, can be written with a new radial variable $y$, which is related to $r$ by 
\be y=r-{r_0^3\over 4r^2}+O(1/r^5) ,\ee 
as
\be \label{metric} ds^2=  dy^2+{y^2\over 36}\Bigl(da^2 +db^2+dc^2 -
{r_0^3\over 2y^3} \left(f_1\,da^2+f_2\,db^2+f_3\,dc^2\right)
+O(r_0^6/y^6)\Bigr),
\ee 
where $r_0$ is a parameter denoting the length scale of $X_i$, and $(f_{i-1},f_i,f_{i+1})=(1,-2,1)$ (indices are understood mod 3). 
When $y\rightarrow \infty$ or $r\rightarrow \infty$, this becomes precisely the cone (\ref{cone}). 

We will need to study the 3-cycles of $Y$ in order to understand the relations between the periods of the $M$-theory $C$-field and the membrane instanton amplitudes, which we shall need in order to describe the moduli space. 

The 3-cycles $D_j$ of $Y$ are given by projections of the $j^{th}$ factor of $SU(2)^3$ to $Y$. Hence, $D_j\cong \S^3$. The third Betti number of $Y$ is two, so the three $D_j$ satisfy the relation
\be \label{cycleconstraint} D_1+D_2+D_3=0 .\ee
The intersection numbers of the $D_i$ are given by 
\be D_i\cdot D_j = \delta _{j,i+1}-\delta _{j,i-1} .\ee
At $X_i$, where the $i^{th}$ factor is filled in, $D_i$ shrinks to zero and the relation (\ref{cycleconstraint}) reduces to $D_{i-1}+D_{i+1}=0$ (where again the indices are understood mod 3). 

At each $X_i$, there is a supersymmetric 3-cycle $Q_i$ given by $g_i=0$. It can be shown that $Q_i$ is homologous to $\pm D_{i-1}$ and $\mp D_{i+1}$, where the sign depends on orientation. 

A manifold still has $G_2$ holonomy up to third order in $r_0 /y$ if we take the $f_j$ of (\ref{metric}) to be any linear combination of $(1,-2,1)$ and its permutations -- so we have $G_2$ holonomy as long as
\be \label{volumes} f_1+f_2+f_3=0 .\ee 

These $f_j$ can be interpreted as volume defects of the cycle $D_j$ at infinity: the volume of $D_j$ depends linearly on a positive multiple of $f_j$. Furthermore, since at the classical manifold $X_i$, only one of the $D_j$ vanishes, only one of the $f_j$ (namely $f_i$) can be negative. So the classical moduli space may contain manifolds with the relation (\ref{volumes}) as long as only one of the $f_j$ is negative \cite{AW,CGLP}.

The periods of the $C$-field along the cycles $D_j$ are $\alpha _j=\int _{D_j}C$. We combine them with the $f_j$ into holomorphic observables $\eta _j$ where now the $C$-field period is a phase:

\be \label{Yeta} \eta _j= \exp{\Bigl ( \frac{2k}{3}f_{j-1}+\frac{k}{3}f_j+i\alpha _j \Bigr )},\ee
where $k$ is a parameter. The relation (\ref{volumes}) means that the $\eta _j$ are not independent, but instead they obey 
\be \eta _1 \eta _2 \eta _3=\exp{\Bigl ( i\sum \alpha _j\Bigr )} .\ee 
(It can be shown that due to a global anomaly in the membrane effective action, the right hand side above is $-1$). 

The moduli space at the classical approximation is given by three branches $\N _i$, each of which contains one of the points $X_i$ with $r_0\rightarrow \infty$. On $X_i$, $\alpha _i$ vanishes and the parameters $f_j$ are such that $\eta _i=1$. So on $\N _i$ the functions $\eta _j$ obey
\be \eta _i ~=~1, \;\; \eta _{i-1}\eta_{i+1}=-1 .\ee

At the quantum level, there are corrections to this statement. It has been suggested in \cite{AMV} that the different classical points $X_i$ are continuously connected to one another. Hence they should appear on the same branch of the moduli space $\N$. We proceed now with the assumption that the only classical points are the $X_i$, which are the points where some of the $\eta _j$ have a zero or pole. As explained in \cite{AW}, since a component of $\N$ which contains a zero of a holomorphic  function $\eta _j$ must also contain its pole, and since the only points at which the $\eta _j$ are singular are associated with one of the $X_i$, it follows indeed that all $X_i$ are contained on a single component of $\N$. Furthermore, each $\eta _j$ has a simple zero and simple pole in $\N$. The existence of such functions on $\N$ means that the branch containing the zero and pole has genus zero. In addition, any of the $\eta _j$ can be identified as a global coordinate of $\N$. Choosing any $\eta _j$ gives a complete description for this branch of $\N$.

\section{Quotients and Low Energy Gauge Groups}\label{quotients}
Here, we begin our study of manifolds which are asymptotic to a cone over quotients of $Y$. We shall consider a discrete group action of $\G= \G_1\times \G_2 \times \G_3$ on $Y$ where the $\G_i$ will be chosen from $ADE$ subgroups of $SU(2)$ in such a way that the low energy physics is known.

We begin with the simplest case where $\Gamma =\Z_p \times \Z_q \times \Z_r$. Each ${\bf Z}_n$ is embedded in $SU(2)$ via
\be \beta ^k= \left (\begin{array}{c c} e^{2\pi ik/n}&0\\0&e^{-2\pi ik/n}\end{array}\right ),
\ee
where $\beta$ is the generator of ${\bf Z}_n$ and $k=0,1, \ldots , n-1$. 
The action of $\G$ on $Y=SU(2)^3/SU(2)$ is given by
\be (\gamma, \delta , \epsilon) \in \Z_p \times {\bf Z}_q \times {\bf Z}_r: (g_1, g_2, g_3)\mapsto (\gamma g_1 , \delta g_2 , \epsilon g_3) ,\ee
and we denote the resulting quotient space by $Y_\Gamma$. 

The spaces $X_{i,\Gamma}$, obtained by filling in the $i^{th}$ $SU(2)$ factor of $Y_\G$, 
are quotients of ${\bf R}^4\times {\bf S}^3$ where the ${\bf R}^4$ corresponds to the filled-in factor. Choosing $i=1$ and gauging $g_2$ away using the right diagonal $SU(2)$ action, the identification corresponding to $(\gamma ^k, \delta ^l, \epsilon ^m)\in \G$ is 
\be \label{zpzqzraction}(g_1, 1, g_3)\sim (\gamma ^k g_1 \delta ^{-l}, 1, \epsilon ^mg_3 \delta ^{-l}) ,\ee
where $g_1\in {\bf R}^4$ and $g_3\in SU(2)\sim \S ^3$. The set $(0,1,g_3)$ with $g_3$ varying in $SU(2)$ is a fixed point of the action of the ${\bf Z}_p$ subgroup of $\G$, and this singularity is identical to the standard $A_{p-1}$ singularity of codimension four of the form $\R ^4/\Z _p$ or $\C ^2 /\Z _p$, which gives an $SU(p)$ gauge theory at low energies. 

Depending on the values of the integers $q$ and $r$, there may be additional, unfamiliar singularities for which we do not know the low energy physics. Namely, there may be values of $g_3$ which are fixed under a non trivial subgroup of $\Z _q\times \Z _r$, i.e. where the following holds 
\be \label{fixed}\epsilon ^mg_3 \delta ^{-l}=g_3 .\ee
This is the same as looking for elements $g_3$ of $SU(2)$ which diagonalize $\delta ^l$: 
\be \label{diagfixed}\epsilon ^m =g_3 \delta ^l g_3^{-1}.\ee
Choosing the orders $q$ and $r$ of $\delta$ and $\epsilon$ to be relatively prime, $(q,r)=1$, ensures that there are no solutions of this equation (since then the orders of the left and right hand sides of (\ref{diagfixed}) are relatively prime). Similarly, we choose $(p,q)=(p,r)=1$, and so there are no singularities at $X_{i,\G}$ other than the $ADE$ singularities whose low energy physics is known: an $A_{p-1}$ singularity on ${\bf S}^3/({\bf Z}_q \times {\bf Z}_r )$ at $X_{1,\Gamma}$, an $A_{q-1}$ singularity on ${\bf S}^3/({\bf Z}_r \times {\bf Z}_p)$ at $X_{2,\Gamma}$, and an $A_{r-1}$ singularity on ${\bf S}^3/({\bf Z}_p \times {\bf Z}_q)$ at $X_{3,\Gamma}$, with the discrete group action on ${\bf S}^3$ given by the appropriate cyclic permutation of the action on $g_3$ in (\ref{zpzqzraction}).

Now consider also the non-abelian $ADE$ groups. Again, we would like to choose $\Gamma$ such that we will only get singularities whose physics at low energies we understand -- namely, $ADE$ singularities. For this purpose we review the relevant properties of the $DE$ groups. For information about these groups, see \cite{Klein}.

As in the abelian case, we let $\G=\G_1\times \G_2 \times \G_3$ act on $Y$ by
\be \label{g1g2g3action}(\gamma, \delta , \epsilon) \in \G_1 \times \G_2 \times \G_3: (g_1, g_2, g_3)\mapsto (\gamma g_1 , \delta g_2 , \epsilon g_3) ,\ee
from which equations (\ref{zpzqzraction}) and (\ref{fixed}) follow in the same way as before.

The binary dihedral groups ${\bf D}_{q}$ have order $4q-8$ and are generated in $SU(2)$ by two elements:
\be {\bf D}_{q}=  \left < \left( \begin{array}{c c} e^{\frac{\pi i}{q-2}}&0\\0&e^{-\frac{\pi i}{q-2}}\end{array}\right), \left( \begin{array}{c c} 0&1\\-1&0\end{array}\right) \right >.
\ee
Since all ${\bf D}_q$ groups share an element of order 4, we cannot choose more than one of the $\G_i$ to be a dihedral group, since otherwise we would get solutions to (\ref{fixed}). Hence we let $\Gamma = {\bf Z}_p \times {\bf D}_q \times {\bf Z}_r$ with $(p,r)=(p,2(q-2))=(r,2(q-2))=1$. 

We turn to the $E$ series. A singularity $\R ^4/G$ which gives at low energies ${\bf E}_6$, ${\bf E}_7$, or ${\bf E}_8$ gauge groups corresponds to $G$ being the tetrahedral group ${\bf T}_{24}$, the octahedral group ${\bf O}_{48}$, or the icosahedral group ${\bf I}_{120}$ . The orders of these groups are 24, 48, and 120 respectively, and each of them has elements of orders 3 and 4, so we cannot have more than one $E$ group appearing in $\Gamma$. The group ${\bf I}_{120}$ also has elements of order 5. Hence, in addition to $(p,r)=1$, for $\Gamma = {\bf Z}_p \times {\bf E}_6 \times {\bf Z}_r$ or $\Gamma = {\bf Z}_p \times {\bf E}_7 \times {\bf Z}_r$, we need also $(p,2\cdot 3)=(r, 2\cdot 3)=1$, and for $\Gamma = {\bf Z}_p \times {\bf E}_8 \times {\bf Z}_r$, we need $(p,2\cdot 3 \cdot 5)=(r, 2\cdot 3\cdot 5)=1$.

Therefore, our group $\G$ is always chosen to be of the form $\G=\Z _p \times \G_2 \times \Z _r$ where $\G _2$ is an $A$, $D$, or $E$ group, and $p$, $r$, and $\G _2$ satisfy the conditions noted above, which can be summarized by 
\be (p,N)=(r,N)=(p,r)=1, \ee
where $N$ is the order of the group $\G _2$. At $X_{1,\G}$ we have an $A_{p-1}$ singularity on ${\bf S}^3/(\G_2 \times {\bf Z}_r )$, at $X_{3,\Gamma}$ we have an $A_{r-1}$ singularity on ${\bf S}^3/({\bf Z}_p \times {\G _2})$, and at $X_{2,\G}$ we have an $A$, $D$, or $E$ singularity on 
${\bf S}^3/({\bf Z}_r \times {\bf Z}_p)$, where here the discrete group action is given by the appropriate cyclic permutation of the action on $g_3$ in (\ref{zpzqzraction}). 

The low energy gauge theories obtained from compactifying $M$-theory on $\R ^4\times X_{i, \G}$ are listed in the following table. Each entry contains the gauge group and the compact 3-manifold which is the locus of the $ADE$ singularity. 

\begin{table}
\be
\label{gts}
\begin{array}{c|c|c|c|} \G_2&X_{1,\G}&X_{2,\G}&X_{3,\G}\\ \hline
\Z_q & SU(p) ~~~ \S^3/(\Z_q\times \Z_r) & SU(q) ~~~ \S^3/(\Z_r \times \Z_p)& SU(r) ~~~ \S^3/(\Z_p\times \Z_q) \\
{\bf D}_q & SU(p) ~~~ \S^3/({\bf D}_q\times \Z_r) & SO(2q)~~~\S^3/(\Z_r \times \Z_p)& SU(r) ~~~ \S^3/(\Z_p\times {\bf D}_q) \\
{\bf T}_{24}&SU(p) ~~~ \S^3/({\bf T}_{24}\times \Z_r) & {\bf E}_6~~~~~~~~\S^3/(\Z_r \times \Z_p) &SU(r) ~~~ \S^3/(\Z_p\times {\bf T}_{24}) \\
{\bf O}_{48}&SU(p) ~~~ \S^3/({\bf O}_{48}\times \Z_r)&{\bf E}_7~~~~~~~~\S^3/(\Z_r \times \Z_p)&SU(r) ~~~ \S^3/(\Z_p\times {\bf O}_{48})\\
{\bf I}_{120}&SU(p) ~~~ \S^3/{(\bf I}_{120}\times \Z_r)&{\bf E}_8~~~~~~~~\S^3/(\Z_r \times \Z_p)&SU(r) ~~~ \S^3/(\Z_p\times {\bf I}_{120})\\ \hline
\end{array}
\ee
\end{table}

As we shall see below, for the cases where $\G _2$ is a $D$ or $E$ group, there are additional semiclassical points where the low energy gauge group is different from those listed above.

We note that for the case with $r=1$, $X_{3,\G}$ is smooth and its low energy theory has no gauge symmetry. If also $p=1$, $X_{1,\G}$ is smooth as well (this is the case studied in \cite{AW}). 
\section{The Curve of M-theories on the Quotient}\label{space}

\subsection{Classical geometry}

The 3-cycles $D_i'$ of $Y_\G$ are the projections of the $i^{th}$ factor of $SU(2)^3$ to $Y_\G$. Hence, for $\G=\Z _p \times {\G _2} \times \Z _r$ we have
\bea D_1'&=&\S ^3/\Z_p ,\\
D_2'&=&\S ^3/{\G _2},\\
D_3'&=&\S ^3/\Z_r .
\eea
Using the relation (\ref{cycleconstraint}) in $Y$ and the fact that $D_1 \in Y$ projects to a $p$-fold cover of $D_1' \in Y_\G$, as well as cyclic permutations of this fact, we find
\be pD_1'+ND_2'+rD_3' =0,\ee
where $N$ is the order of the group ${\G_2}$. To study the intersection numbers of the $D_i'$ we note that $D_1'\in Y_\G$ lifts to $NrD_1\in Y$, and similar statements are true for the other $D_i'$. Counting the intersection numbers in $Y$ and then dividing by $pNr$ (since there are $pNr$ points in $Y$ which project to one point in $Y_\G$), we get
\be D_1'\cdot D_2'=r ,\hskip 1cm D_2'\cdot D_3' = p ,\hskip 1cm D_3'\cdot D_1' =N .\ee
Here we see that the $D_i'$ generate the third homology group of $Y_\G$: since $(r,N)=1$, we can find integers $m,n$ such that
\be D_1'\cdot (mD_2'+nD_3')=mr-nN=1 ,\ee
and similarly for the other cycles.

We define the periods of the $M$-theory $C$-field at infinity by 

\be \alpha_j'=\int_{D_j'}C~~{\rm mod}~2\pi . \ee
Note that these are related to the $\alpha _j$ of $Y$ by 
\be \label{alpha'alpha} \alpha _1=p\alpha _1' ,\hskip 1cm \alpha _2=N\alpha _2' ,\hskip 1cm \alpha _3=r\alpha _3' .\ee


\subsection{Classical moduli space}

We define our holomorphic observables to be the following functions of the periods $\alpha _j'$ and of the volumes $f_j$:
\begin{eqnarray} \eta _1&=&\exp {\Bigl (\frac{2k}{3p}f_3+ \frac{k}{3p}f_1+i\alpha _1 ' \Bigr )} ,\nonumber \\ 
\eta _2&=&\exp {\Bigl ( \frac{2k}{3N}f_1+ \frac{k}{3N}f_2+i\alpha _2 ',\Bigr )} ,\nonumber \\
\eta _3&=&\exp {\Bigl ( \frac{2k}{3r}f_2+ \frac{k}{3r}f_3+i\alpha _3 '. \Bigr )} . \nonumber
\end{eqnarray}
These functions are adopted from (\ref{Yeta}), where we substitute the expressions in (\ref{alpha'alpha}) for the periods and then take the largest possible root that still leaves the $\eta _i$ invariant under $\alpha _j'\mapsto \alpha _j'+2\pi$.

The periods of the $C$-field are interpreted as the phases of the holomorphic observables.  

Due to (\ref{volumes}), we have
\be \eta _1 ^p\eta _2 ^N\eta _3 ^r=\exp{\Bigl ( i\sum _j \alpha _j'\Bigr )} .\ee

The $\eta _j$ have zeros or poles at the semiclassical points $X_{i, \G}$  with large $r_0$ in which the $f_j$ diverge. As in Section \ref{Y}, classically at the point $X_{1,\G}$, $\eta _1=1$ and $\alpha _1'=0$. Hence, at this point
\be \label{classicalmoduli}\eta _2 ^N\eta _3^r=\exp{\Bigl ( i(\alpha _2'+\alpha _3')\Bigr )} , \ee
so when $\eta _2$ has a pole, $\eta _3$ has a zero and vice versa. In fact, the order of the zeros or poles of $\eta _2$ must be a multiple of $r$, and similarly the order of the zeros or poles of $\eta _3$ must be a multiple of $N$ for this equation to hold. In the classical approximation, there are three branches $\N _i$ of the moduli space, on which we have $\eta _i =1$ and $\eta _{i\pm 1}$ obeying the relation (\ref{classicalmoduli}) for $i=1$ or cyclic permutations of it for $i=2,3$.

\subsection{Quantum curve via membrane instantons}
To study the quantum curve, we study the singularities, i.e. the zeros and poles of the holomorphic observables $\eta _j$, which correspond to the classical points $X_{i,\G}$ with $r_0\rightarrow \infty$. We shall use a relation between  the $\eta _j$ and the amplitude for membrane instantons which wrap on supersymmetric cycles $Q$ in $X$. Using chiral symmetry breaking of the low energy gauge theories, we find a clear relation between the local parameter on the moduli space and our observables, and hence can describe the moduli space.

A supersymmetric cycle in $X_{i,\Gamma}$ is given by the 3-manifolds $Q_i$ given by $g_i=0$:
\bea Q_1&=&\S ^3/({\G _2} \times \Z _r) ,\\
Q_2&=&\S ^3/(\Z _r \times \Z _p) ,\\
Q_3&=&\S ^3/(\Z _p \times {\G _2}). 
\eea
At $X_{1,\G}$, $Q_1$ is homologous (up to orientation) to the $D_j'$ as follows:
\bea \label{homologous} rQ_1\sim D_2' ,\\ NQ_1\sim D_3' ,\eea
and cyclic permutations of that give the relations at $X_{2,\G}$ to be $pQ_2\sim D_3'$ and $rQ_2\sim D_1'$, and at $X_{3,\G}$ we have $NQ_3\sim D_1'$ and $pQ_3\sim D_2'$.

We now study the zeros and poles of the $\eta _j$. To understand the orders of the zeros and poles, we must compare the $\eta _j$ to the true local parameter on ${\cal N}_\G$ around each $X_{i,\G}$ with large $r_0$. 

One would expect at first that the membrane instanton amplitude $u$ itself, given near $X_{i,\G}$ by 
\be u=\exp{\Bigl (-TV(Q_i)+i\int _{Q_i}~C~ \Bigr )},\ee 
where $T$ is the membrane tension and $V(Q_i)$ is the volume of $Q_i$, would be a good local parameter near $X_{i,\G}$. However, at low energies we have a supersymmetric $A$, $D$, or $E$ gauge theory in four dimensions, and due to chiral symmetry breaking, we expect the good local parameter -- the gluino condensate -- to be $u^{1/h}$ where $h$ is the dual Coxeter number of the gauge group.

We now compare phases of the $\eta _j$ to the phase of $u$. Let $P_{i,\G}$ correspond to the manifolds $X_{i,\G}$ with large $r_0$. For the case where ${\G _2}=\Z_q$, at $P_{1,\G}$ equation (\ref{homologous}) implies that the phase $\int _{D_2'}C$ of $\eta _2$ is related to the phase $\int _{Q_1}C$ of $u$ by $\int _{D_2'}C\sim r\int _{Q_1}C$. Since the good local parameter is actually $u^{1/p}$ due to chiral symmetry breaking of the $SU(p)$ gauge theory at $P_{1,\G}$, the true order of the zero of $\eta _2$ at $P_{1,\G}$ is $pr$. The same calculation for the other $\eta _j$ and $P_{i,\G}$ gives the orders of zeros and poles shown in the following table:

\be \begin{array}{c|c |c| c}{\G_2}=\Z_q & P_{1, \Gamma} & P_{2, \Gamma} & P_{3, \Gamma}  \\ \hline \eta _1 & 1 & \infty ^{qr} & 0^{qr} \\ \hline \eta _2 &0^{pr} &1&\infty ^{pr} \\ \hline \eta _3 & \infty ^{pq} & 0^{pq} & 1 
\end{array}
\ee
The cases where ${\G_2}$ is a $D$ or $E$ group give similar tables, except that in these cases we get extra semiclassical points in the same way as in \cite{AW}: for the case ${\G_2}={\bf D}_q$, we have 

\be \begin{array}{c|c |c| c|c } {\G_2}={\bf D}_q& P_{1, \Gamma} & P_{2, \Gamma} &  P_{2', \Gamma } & P_{3, \Gamma}   
\\ \hline \eta _1 & 1 & \infty ^{rh} & \infty ^{2rh'}& 0^{rN} 
\\ \hline \eta _2 &0^{rp} &1& -1 & \infty ^{rp} 
\\ \hline \eta _3 & \infty ^{Np} & 0^{hp} & 0^{2h'p}& 1 
\end{array}\; \; ,
\ee
where $h=2q-2$, $h'=q-3$, and $h+2h'=N$. The low energy gauge theory at $P_{2',\G}$ has gauge group $Sp(q-4)$. 

For ${\G _2}$ in the $E$ series, the table is
\be \label{Etable} \begin{array}{c|c |c| c|c }{\G_2}={\bf E}_a & P_{1, \Gamma} & P_{2, \Gamma} &  P_{\mu  t, \G } & P_{3, \Gamma}   
\\ \hline \eta _1 & 1 & \infty ^{rh} & \infty ^{ rth_t}& 0^{rN} 
\\ \hline \eta _2 &0^{rp} &1& e^{2\pi i\mu/t} & \infty ^{rp} 
\\ \hline \eta _3 & \infty ^{Np} & 0^{hp} & 0^{pth_t}& 1 
\end{array}\; \; ,
\ee
where $t,h_t, \mu$ are given for each ${\bf E}_a$ as follows: let $k_i$ be the Dynkin indices of ${\bf E}_a$, and let $t$ be the positive integers which divide some of the $k_i$; $\mu$ runs over positive integers less than $t$ that are prime to $t$, unless $t=1$ in which case $\mu =0$; $h_t$ is the dual Coxeter number of the associated group $K_t$ whose Dynkin indices are $k_i/t$ where here the $k_i$ run through the indices of ${\bf E}_a$ that divide $t$. The $t$ and $h_t$ obey the relation $\sum th_t=N$. The low energy gauge group is given by the $ADE$ group corresponding to $K_t$.  

From the relation $\sum th_t=N$ and the tables above, we see that for each $\eta _j$, the total number of zeros and poles is equal. Since the total number of zeros is the same as the total number of poles for each of the $\eta _j$, it seems reasonable to assume that we have found all the zeros and poles, and hence all the semiclassical limits in our moduli space. It would seem, therefore, that we can now proceed to describe the moduli space completely, by writing our functions $\eta _i$ explicitly and identifying the points $P_{i,\G}$ with values of a good coordinate on the moduli space. However, as we shall see, we run into a few puzzles.

The first question we ask is: what can be said about the genus of $\N _\G$? For the cases $p=r=1$, which are the cases considered in \cite{AW}, the function $\eta _2$ has a simple zero and a simple pole, and hence can be identified with a global coordinate on the moduli space, which can then be claimed to have genus zero. If $p,r>1$, this is not so: none of our $\eta_j$ have just a simple zero and pole, so we cannot identify the moduli space with any of the $\eta _j$, and we do not know the genus. 

However, the simplest result would be that the curve has genus zero, and we proceed with this assumption. Hence, we assign the curve a global coordinate $z$, write the $\eta_j$ as holomorphic functions of $z$, and see how well we can describe the curve.

For the case ${\G_2}=\Z_q$, this turns out to be straightforward; we may fix $P_{1,\G}$ at $z=0$, $P_{2,\G}$ at $z=1$, and $P_{3,\G}$ at $z=\infty$, and then write our functions:
\bea \eta _1&=&\frac{1}{(1-z)^{qr}} ,\\
\eta _2 &=& z^{pr},\\
\eta _3&=& \frac{(1-z)^{pq}}{z^{pq}}.
\eea 
This description is unique up to possible overall factors which are related to an anomaly in the membrane effective action, analogous to the one described in Section~5 of \cite{AW}. 

For ${\G_2}={\bf D}_q$, we run into a puzzle. Once we fix the first three points, we have to find at what value $z_4$ the fourth point $P_{2',\G}$ sits: our functions in this case are 
\bea \eta _1&=& \frac{z_4^{2rh'}}{(1-z)^{rh}(z_4-z)^{2rh'}} ,  \\
\eta _2&=& z^{rp}, \\
\eta _3&=& \frac{(1-z)^{ph}(z_4-z)^{2h'p}}{z^{Np}},
\eea
again up to overall factors. The forms of $\eta _1$ and $\eta _3$ do not constrain $z_4$, but to satisfy $\eta _2(z_4)=-1$, we need $z_4^{pr}=-1$ for which there are $pr$ solutions. A similar situation arises for ${\G _2}$ in the $E$ series, where there are $pr$ choices for each point beyond the first three. 

The $pr$ solutions, however, should correspond to the same point in the moduli space of $M$-theories, since they correspond to the same theory. Hence, it seems that we have a redundancy in our description of the moduli space; we should impose a symmetry on $\N _\G$ which identifies the different values of $z_4$. 

There is another, more serious puzzle which shows up, also involving possible extra classical points on $\N _\G$: from table (\ref{gts}), we see that our low energy gauge theory is compactified on a manifold which is not simply connected, but rather is of the form $\S^3/H$ for some discrete group $H$. Hence its fundamental group is equal to $H$. Therefore, it is possible to construct theories which have gauge fields with non-trivial Wilson loops which break the gauge symmetry. Where in $\N _\G$ do these theories lie?

For the case ${\G_2}=\Z_q$, the point $P_{1,\G}$ can have Wilson loops which are conjugacy classes of elements of $SU(p)$ of order $qr$. One can show that, when $p,q,r$ are relatively prime, the number of inequivalent such elements is 
\be \frac{1}{p}\left ( \begin{array}{c} p+qr-1\\qr-1 \end{array} \right )=\frac{(p+qr-1)!}{p!(qr)!} \ee 
with cyclic permutations for $P_{2,\G}$ and $P_{3,\G}$. Furthermore, for Wilson loops that break $SU(p)$ in a way that leaves $s-1$ factors of $U(1)$, i.e. 
\[ SU(p)\longrightarrow \Pi _{i=1}^s SU(n_1)\times U(1)^{s-1}, \] 
where $\sum n_i=p$, the number of inequivalent Wilson loops is 
\be \label{snumber}\frac{s}{pqr}\left ( \begin{array}{c} p\\s \end{array} \right ) \left ( \begin{array}{c} qr\\s \end{array} \right ).
\ee 
Each set of theories with a given number $s-1$ of $U(1)$ factors should lie on a separate component $\N _{s,\G}$ of the moduli space, since smooth interpolation means that the number of massless modes -- which corresponds to $U(1)$ fields -- is constant on each component. For $s>1$, we know that the theories on $\N _{s,\G}$ do not have a mass gap due to the massless $U(1)$ field. On the other hand, the theories corresponding to the points $P_{i,\G}$ with no non trivial Wilson loops are believed to have a mass gap. Hence we claim that $\N_{1,\G}$ contains theories with a mass gap. 

For the case $r=1$, we obtain no singularity at $X_{3,\G}$. Hence, for that case the mass gap of the theory at $X_{3,\G}$ means that all of $\N _{1,\G}$ has a mass gap. 

Continuing with the case where $r=1$, we note a manifest symmetry between $p$ and $q$ in the expression (\ref{snumber}) for the number of possible Wilson loops at each level $s$. At first sight, this could support the assertion that these points lie on their own branch of the moduli space, which will interpolate smoothly among them and contain no other singular points. However, chiral symmetry breaking means that the number of vacua at each classical point is given by $\Pi n_i$ which is clearly not symmetric between $p$ and $q$, and spoils the counting of the orders of zeros and poles.

Going back to general $r$ and looking at $\N_{1,\G}$ only, we see that we have smooth interpolation among theories with different gauge groups: $SU(p)$, $SU(q)$, and $SU(r)$ when ${\G_2} =\Z_q$; $SU(p)$, $SO(2q)$, $Sp(q-4)$, and $SU(r)$ when ${\G_2} ={\bf D} _q$; and analogously for ${\G_2}$ in the $E$ series, where we interpolate between $SU(p)$, $K_t$, and $SU(r)$, with $K_t$ as described after table (\ref{Etable}). Similarly, the other branches $\N _{s,\G}$ smoothly interpolate among theories with these gauge groups broken by Wilson lines.

\vskip 1cm
\centerline{{\bf Acknowledgements}}
\no The author is very grateful to Edward Witten for useful and inspiring discussions and for guidance at every stage of this project. This research has been supported in part by a Paul and Daisy Soros Fellowship for New Americans and in part by a National Science Foundation Graduate Research Fellowship.

\vskip 2cm

\end{document}